%% file: main.tex
\documentclass[fleqn]{2023SCGE}
\usepackage{amsmath}
\usepackage{appendix}
\usepackage{graphicx}
\usepackage{seqsplit}
\usepackage{natbib}
\usepackage{aas_macros}

\newcommand{\grays}{$\gamma$-rays\ }

\begin{document}

\ensubject{subject}

\ArticleType{Article}
\SpecialTopic{SPECIAL TOPIC: }
\Year{2023}
\Month{January}
\Vol{66}
\No{1}
\DOI{??}
\ArtNo{000000}
\ReceiveDate{ }
\AcceptDate{ }

\title{Deep view of Composite SNR CTA1 with LHAASO in $\gamma$-rays up to 300 TeV}

\author{LHAASO Collaboration\footnote{Corresponding author:xisq@ihep.ac.cn, liyz@ihep.ac.cn,bli@smail.nju.edu.cn,huangyong96@ihep.ac.cn,chensz@ihep.ac.cn} \\(The LHAASO Collaboration authors and affiliations are listed after the references.)}{}%

\AuthorMark{Zhen Cao}
\AuthorCitation{Zhen Cao, et al}

\abstract{
The ultra-high-energy (UHE) gamma-ray source 1LHAASO J0007+7303u is positionally associated with the composite SNR CTA1 that is located at high Galactic Latitude $b\approx 10.5^\circ$. This provides a rare opportunity to spatially resolve the component of the pulsar wind nebula (PWN) and supernova remnant (SNR) at UHE. This paper conducted  a dedicated data analysis of 1LHAASO J0007+7303u using the data collected from December 2019 to July 2023. This source is well detected with significances of 21$\sigma$ and 17$\sigma$ at 8$-$100 TeV and  $>$100 TeV, respectively.  The corresponding extensions are determined to be 0.23$^{\circ}\pm$0.03$^{\circ}$ and 0.17$^{\circ}\pm$0.03$^{\circ}$. The emission is proposed to originate from the relativistic electrons accelerated within the PWN of PSR J0007+7303. The energy spectrum is well described by a power-law with an exponential cutoff function $dN/dE = (42.4\pm4.1)(\frac{E}{20\rm\ TeV})^{-2.31\pm0.11}\exp(-\frac{E}{110\pm25\rm\ TeV})$ $\rm\ TeV^{-1}\ cm^{-2}\ s^{-1}$in the energy range from  8 TeV to  300 TeV, implying a steady-state parent electron spectrum $dN_e/dE_e\propto (\frac{E_e}{100\rm\ TeV})^{-3.13\pm0.16}\exp[(\frac{-E_e}{373\pm70\rm\ TeV})^2]$ at energies above $\approx 50 \rm\ TeV$.  The cutoff energy of the electron spectrum is roughly equal to the expected current maximum energy of particles accelerated at the PWN terminal shock. Combining the X-ray and gamma-ray emission, the current space-averaged magnetic field can be limited to $\approx 4.5\rm\ \mu G$. To satisfy the multi-wavelength spectrum and the $\gamma$-ray extensions, the transport of relativistic particles within the PWN is likely dominated by the advection process under the free-expansion phase assumption.
}



\maketitle


\section{Introduction}

CTA1 is a famous composite supernova remnant (SNR), characterized by a shell SNR and a central pulsar wind nebula (PWN).
It was discovered in the radio band and first proposed as a SNR by \cite{1960PASP...72..237H}. 
The detailed radio observation further revealed a northwest-incomplete shell with a diameter $\approx 1.8^{\circ}$ and a bridge-like structure towards the center, with a kinematic distance of $\approx 1.4$ kpc according to the associated $H_{\uppercase\expandafter{\romannumeral1}}$ shell \citep{1993AJ....105.1060P,1997A&A...324.1152P,2011A&A...535A..64S}.
The first X-ray detection in the direction of CTA1 was made by ROSAT\cite{1995ApJ...453..284S}, which revealed both thermal emission from the outer shock-heated region and a non-thermal component from the central region. Subsequent studies of the central X-ray source (RX J0007.0+7302) with ASCA, XMM-Newton, Chandra and Suzaku observatory have resolved the X-ray emission into a point-like source likely corresponding to the central pulsar, and a diffuse nebula, implying a X-ray PWNe driven by the central active pulsar \citep{1997ApJ...485..221S,2004ApJ...612..398H,2004ApJ...601.1045S,2012MNRAS.426.2283L}. 

The point-like source has been firmly identified as a pulsar (PSR J0007+7303), owing to the successful search for pulsation in GeV band \citep{2008Sci...322.1218A} and X-ray band \citep{2010ApJ...725L...1L,2010ApJ...725L...6C}. PSR J0007+7303, with a period of $\approx 316$ ms, has a sufficient spin-down power $\dot{E}= 4.5 \times 10^{35}$ erg~s$^{-1}$ to drive the X-ray PWN. In addition, the corresponding characteristic age $\tau_{c}$ is $\approx 13.9$ kyr,consistent with the SNR age derived by \cite{2004ApJ...601.1045S}. In GeV regime, searches for the extended emission associated with radio shell and/or X-ray nebulae are ongoing 
\citep{2012ApJ...744..146A,2013ApJ...773...77A,2016ApJ...831...19L}. The latest results of the $E>1$ GeV extended emission searching shows a $ \approx 0.98^{\circ}$ disk-shaped source, potentially associated with the SNR due to its overlap with the radio emission contours \citep{2018ApJS..237...32A}. In the  $E > 50$ GeV energy range, \cite{2024MNRAS.tmp..747Z} report a possible  $\sim 0.4^\circ$ extended $\gamma$-ray emission from the PWN.
 
Composite SNR CTA1 is associated with an extended TeV $\gamma$-ray source, VER J0006+729, discovered by VERITAS experiment in the energy range $0.6\rm\ TeV$$-$$17.8 \rm\ TeV$ \citep{2013ApJ...764...38A}. The TeV source  VER J0006+729 shows an elliptical Gaussian morphology with 1$\sigma$ angular extension of $ 0.30^{\circ}$ along the major axis and $ 0.24^{\circ}$ along the minor axis, with the orientation of the major axis of $\approx 17.8^{\circ}$ west of north, centered at the position of the pulsar PSR J0007+7303.  
Due to the positional and morphological coincidence of the TeV emission with the X-ray PWN, the TeV source VER J0006+729 is discussed as  TeV PWN in studies such as \cite{2013ApJ...764...38A} and \cite{2014JHEAp...1...31T}. Generally, PWNe are clouds of magnetised plasma that are created inside SNRs by the highly relativistic outflow (``wind") of a pulsar,  observed via synchrotron emission produced when energetic electrons interact with the magnetic field, or through inverse Compton (IC) radiation generated by the scattering of electrons off the background photon fields (see \cite{2006ARA&A..44...17G} and \cite{2022hxga.book...61M}, for a comprehensive review). The latter mechanism is believed to be relevant to the TeV $\gamma$-ray emission of the PWN. Alternatively, hadronic mechanisms could also be the origin of TeV emission, although unidentified by observation, in which case relativistic hadrons collide with the ambient medium, producing TeV emission through $\pi_{0}$ decay. The study of UHE $\gamma$-ray emission is crucial in determining whether the $\gamma$-ray emission is predominantly from hadronic processes or the IC process, as IC emission above 100 TeV energies undergoes suppression due to the Klein-Nishina effect.
 
Indeed, the vast majority of Galactic TeV emittiers have been identified in the population of PWNe \citep{2018A&A...612A...2H}. Among them, CTA1 holds a special status as it is located at a relatively high Galactic Latitude $b\approx 10.5^\circ$, experiencing minimal influence from Galactic diffuse emission and nearby TeV sources. This positioning makes it an excellent candidate for accurate measuring characteristics of TeV emission, serving as a valuable object for probing the physical model of TeV PWNe and distinguishing between leptonic and hadronic processes. Thanks to the wide field view, high sensitivity, and broad energy range of the Large High Altitude Air Shower Observatory (LHAASO) \citep{2010ChPhC..34..249C}, our understanding of CTA1 has made impressive progress. Using approximately 2 years of data, the LHAASO collaboration has reported an extended TeV source, 1LHAASO J0007+7303u, which is tentatively shaped by a Gaussian with 1$\sigma$ angular extension of $\sim 0.2^\circ$ at energies $E > 25\rm\ TeV$ and is positionally coincident with  the X-ray/TeV PWN in Composite SNR CTA1.  Particularly noteworthy is the significant detection of the ultra-high-energy (UHE, $E > 100\rm\ TeV$) $\gamma$-ray emission from the source 1LHAASO J0007+7303u, at a 13$\sigma$ confidence level~\citep{2023arXiv230517030C}. 

This paper presents a deep observation of CTA1 with LHAASO using approximately 3 years of data, and proposes some possible mechanisms for the UHE emission through dedicated data analysis and discussion. Section 2 provides a brief introduction to the LHAASO detector array and the analysis method. In Section 3, we report the analysis results based on LHAASO data. A discussion of the multi-wavelength observation and a model for CTA1 is presented in Section 4. Finally, we summarize the conclusions in Section 5.

\section{LHAASO Data Analysis}

LHAASO,constructed on Mountain Haizi, in Sichuan province, China, is a complex extensive air shower (EAS) array with a high sensitivity ($\sim 1\%$ CU 1 year) and a  wide field of view (FOV,$\sim$2.24 Sr for the maximum zenith angle of $50^\circ$) for CRs and $\gamma$-rays. It consists of three subarrays, i.e., Water Cherenkov Detector Array (WCDA), Kilomiter Square Array (KM2A) and Wide Field-of-view Cherenkov Telescope Array (WFCTA). By Combining WCDA and KM2A, LHAASO can cover the energy range from $\sim$1 TeV to $>$ 1 PeV for $\gamma$-ray observation, with the low-energy threshold depending on the zenith angle distribution to some extent. The performance of WCDA and KM2A has been studied in detail employing the Monte Carlo simulations \citep{2024RDTM..tmp...28C}, and calibrated using the measurements of Crab Nebula as a standard candle \citep{2021ChPhC..45b5002A,2021ChPhC..45h5002A}. 

\subsection{Data selection and binning}
The KM2A events utilized in this study were taken during the period from 17th December 2019 to 31st July 2023. The data quality control system and the long-term performance monitoring of KM2A data can be found in \citep{2024arXiv240511826C}.  After rigorous data selection, the total live time amounts to 1216 days. These events were further reduced and reconstructed, according to the selection criteria and reconstructing methods outlined in \cite{2021ChPhC..45b5002A}. We applied a $\gamma$-ray/background discrimination cut to select out all $\gamma$-like events,  with the survival fraction of cosmic ray background events at approximately $4\times10^{-4}$ at 50 TeV energy compared to the CR observation. For the analysis presented in this paper, only KM2A $\gamma$-like events with zenith angles less than 50 degrees and reconstructed energies ($E_{rec}$) above 40 TeV were included. The low-energy threshold is dictated by the higher threshold for the higher zenith angle of the source. At the LHAASO site, the CTA1  (Decl. = 73$^{\circ}$) culminates at a zenith angle $\theta$ = 43.7$^{\circ}$  and lies at zenith angles $\theta < 50 ^{\circ}$ for 6.2 hours per sidereal day. 
We binned our selected data into 5 logarithmically spaced $E_{rec}$ bins per decade. Within each $E_{rec}$ bin, we generated an ``on map" by filling events into a grid of the spatial pixels with dimensions of $\Delta\rm R.A.\times \Delta\rm Decl.=0.1^\circ\times 0.1^\circ$ based on the reconstructed direction. The number of isotropic background events, still predominantly cosmic-ray events, in each spatial pixel was estimated using the ``direct integral method" \citep{2004ApJ...603..355F}, from which we derived the ``background map". 

The WCDA data used in this study were collected from 5th March 2021 to 31st July 2023, covering a total of 796 days of live time. To obtain the available $\gamma$-like event set for source analysis, we implemented quality cuts, direction reconstruction, and  $\gamma$-ray/background discrimination, as detailed in \cite{2021ChPhC..45h5002A}. The number of hit ($N_{hit}$) for each event was selected as a shower energy estimator. We divided our data into 5 intervals of $N_{hit}$, namely [100,200], [200,300], [300,500], [500,800], and [800, 2000]. Within each $N_{hit}$ bin, we  generated the ``on map" and ``backgound map" following the aforementioned procedure for KM2A data. 

A region of interest (ROI) specific to the CTA1 was defined as a  $6^\circ\times 6^\circ$ square region centered at the position of pulsar PSR J0007+7303 (R.A.=1.757$^\circ$; Decl.=73.052$^\circ$). We selected the ``on map" and ``background map" within ROI and  for all energy bins to following analysis. In total, there are 14 energy bins represented by $N_{hit}$ bins for WCDA data  and $E_{rec}$ bins for KM2A data. The angular and energy resolution are dependent on the zenith angle and the energy.  Considering  the position of pulsar PSR J0007+7303, the distribution of zenith angle is constant, and thus the median energy in each energy bin slightly varies with the spectral shape of the source.


\subsection{Maximum Likelihood Analysis and Statistic Test} 
We utilized a maximum likelihood fit using parametric spatial and spectral models to determine the statistical significance of $\gamma$-ray emissions detected by LHAASO.
This process involved generating expected source counts map (``src map") of $\gamma$-ray signals in our ROI through forward-folded models with the detector response. We then compared the sum of the model counts listed in ``src map" and the background events listed in ``background map" with the actual observed counts ($N_{obs}$) listed in ``on map". Given a model $\theta$ with spatial and spectral parameters, we maximized the likelihood of the model comparing with $N_{obs}$ as follows:
 
 \begin{equation}
     ln\mathcal{L}(N_{obs}|\theta) = \sum_{j=0}^{N_{bins}}lnP(N_{obs}^{j}|\theta)
 \end{equation}
where $N_{bins}$ is the number of the data bins in the ROI and the analysis energy, P is the Poisson probability of detecting $N_{obs}^{j}$ events in each bin given the model parameters $\theta$. 

The ratio of the maximum likelihood defines a test statistic (TS):

 \begin{equation}
     {\rm TS} = 2 \ln (\mathcal{L} (\theta_1)/ \mathcal{L}(\theta_0))
 \end{equation}
 where $\theta_0$ and $\theta_1$ represent the parameters of the model of the null and alternative hypotheses, respectively. Based on Wilks' Theorem, the TS follows a $\chi^{2}_{n}$ distribution, where n is the degrees of freedom (dof.) derived from the difference in the number of free parameters between the models. Typically, detection significance is determined by comparing the likelihood of the background-only model (null hypothesis) with that of the signal plus background model (alternative hypothesis).  Unless otherwise mentioned, the TS utilized in this study is for detection significance.  
 We employed the difference in detection significance ($\Delta\rm TS$) to compare the spatial and/or spectral models of the source. It is important to note that the difference in TS cannot be used to quantitatively determine the preferred model when the models are not nested.Alternatively, the Akaike information criterion test \citep[AIC,][]{1974ITAC...19..716A} can be considered. The AIC is defined as AIC = $2k - 2 ln\mathcal{L}$, where $k$ is the number of parameters in the model. In this context, the best hypothesis is considered to be the one that minimizes the AIC.  A qualitative strength of evidence rules to assess the significance of a model is based on  the difference in AIC ($\Delta$AIC) values between the two models. If $\Delta\rm\  AIC > 5$, then it is considered strong evidence against the model with higher AIC and 
$\Delta\rm\ AIC > 10$ is considered as decisive evidence against the model with higher AIC \citep{2007MNRAS.377L..74L,2017Ap&SS.362...70K}.

\section{Results}

  \subsection{Morphology}

We evaluated the TS value for a point-like signal with a power-law spectral shape $dN/dE\sim E^{-2.7}$ at each pixel in the map. In this case,  the significance is represented by $\sqrt{\rm TS}$ . The  $\sqrt{\rm TS}$ map is displayed within a $2.5^\circ\times 2.5^\circ$ square area centered at the position of pulsar PSR J0007+7303 in the energy range of $8{\rm\ TeV} < E < 100\rm\ TeV$ and $E > 100\rm\ TeV$, respectively, as illustrated in Figure 1.  
In both the $8 {\rm\ TeV} < E < 100\rm\ TeV$ and  $E > 100\rm\ TeV$ energy ranges, an obvious $\gamma$-ray emission excess structure is  observed in the inner region of the radio shell. 
\begin{figure}[ht]
\centering
\includegraphics[width=0.9\textwidth]{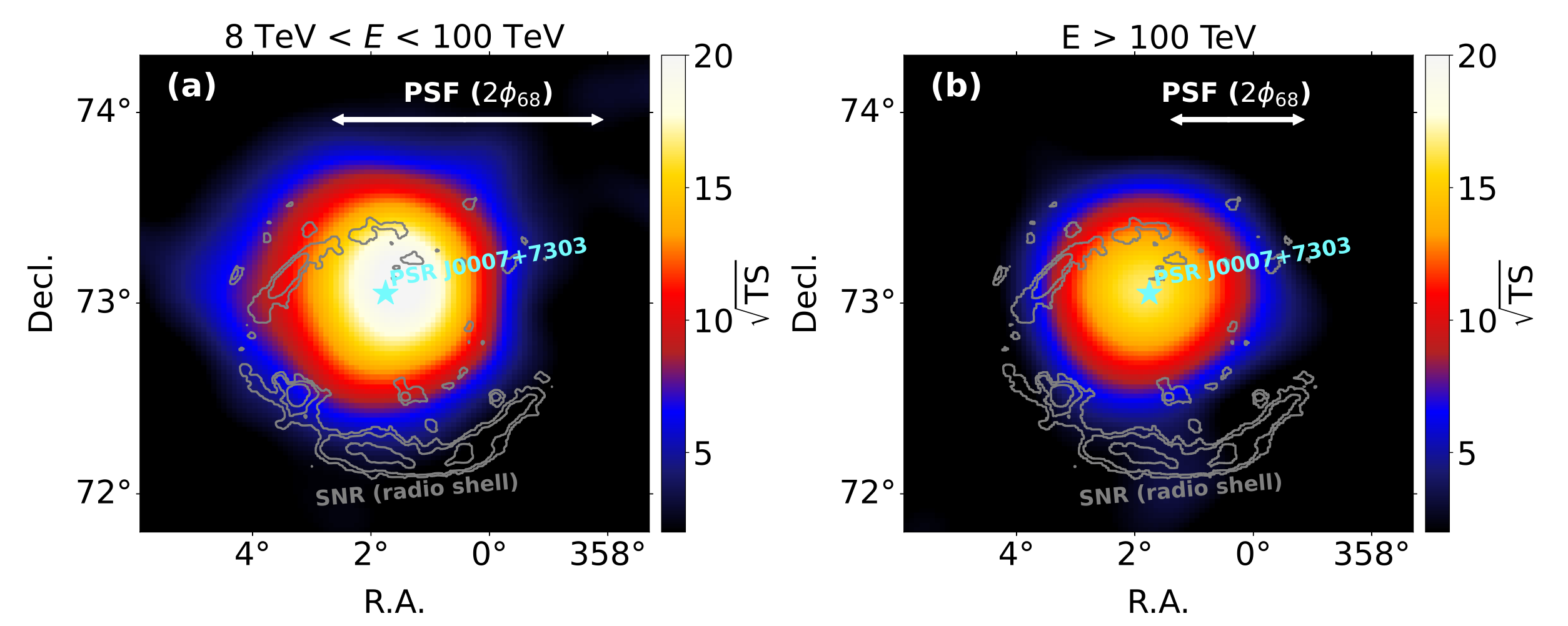}%
\caption{The significance maps of CTA1 in the energy ranges of $8{\rm\ TeV}<E < 100\rm\ TeV$ (a) and $E > 100\rm\ TeV$ (b). Overlaid are the GB6 image countour (4850MHz) in gray. Cyan stars represent the position of the pulsar PSR J0007+7303. }
\label{fig:tsmap1}
\end{figure}

We studied the morphology of the $\gamma$-ray emission in the energy range of $8{\rm\ TeV} < E < 100\rm\ TeV$  and  $E > 100\rm\ TeV$, respectively. 
Due to the impact of angular resolution,  the actual morphology of the $\gamma$-ray emission in  CTA1 should be addressed by using a forward-folded approach. We first explored the routine and mathematical-empirical geometrical models, such as point, Gaussian and disk.  Motivated by the observations of VERITAS in the energy range of $0.6 {\rm\ TeV} < E < 17.7 \rm\ TeV$, we also examined the elliptical Gaussian model. However, due to the limited angular resolution of LHAASO, which is 3-10 times larger than that of Image Atmosphere Cherenkove Telescope (IACT), we encountered convergence issues in the case of all parameters being free. To address this, we maintained the rotation angle fixed. For each energy range, the spectrum of the $\gamma$-ray emission in CTA1 region was initially modeled by a simple power-law for the 3D maximum likelihood analysis. Fit results are detailed in Table~\ref{tab:SFR}.
 
\begin{table*}[h!]
 \centering
 \setlength{\tabcolsep}{0.08in}
 \caption{Morphological Fit Results for CTA1.}
 \label{tab:SFR}
 \begin{tabular}{llccccccccc}
 \hline
 \hline
Energy & Spatial Model & TS & R.A. & Decl.  & $p_{95}$ & Major Axis & Minor Axis& Position Angle& AIC  \\
            &                        &       & deg  & deg    & deg          & deg            & deg          &      deg               & \\
\hline
$>$ 100 TeV & Point & 305.8 & 1.83 & 73.07 & 0.07 & - & - & - & 0.0\\
& Disk & 318.6 & 1.77 & 73.06 & 0.07 & 0.31$\pm$0.07 & - & - & -10.8\\
& Gaussian & 319.0 & 1.77 & 73.06 & 0.08 & 0.17$\pm$0.03 & - & - & -11.2\\

& Elliptical gaussian & 319.2 & 1.78 & 73.06 & 0.08 & 0.18$\pm$0.04 & 0.16$\pm$0.06 & 17.8 (fixed) & -9.2\\
\hline
8 -100 TeV & Point & 431.4 & 1.51 & 73.11 & 0.07 & - & - & - & 0.0\\
& Disk & 474.8 & 1.65 & 73.08 & 0.06 & 0.39$\pm$0.04 & - & - & -41.4\\
& Gaussian & 476.8 & 1.66 & 73.08 & 0.06 & 0.23$\pm$0.03 & - & - & -43.4\\
& Elliptical gaussian & 479.6 & 1.62 & 73.09 & 0.07 & 0.26$\pm$0.07 & 0.18$\pm$0.05 & 17.8 (fixed) & -43.4\\
\hline
\hline
\end{tabular}  
\begin{tablenotes}
\item {$P_{95}$ is the statistical positional uncertainty at a 95\% confidence level. The Major Axis is the radius of 100\% flux region for Disk model and the radius of 39\% flux region for Gaussian model. The AIC values are subtracted from that of the point model for a clear comparison. For the elliptical Gaussian model, we fixed the rotation angle (from west to north 17.8$^\circ$) determined from VERITAS observations to ensure fitting convergence}.
\end{tablenotes}
\end{table*}

Comparing the Gaussian model with the point model, we find that the Gaussian model  improved the fit with a significance of 
6.7$\sigma$ ($\Delta\rm TS= 45.4$)  at energies 8-100 TeV and  3.5$\sigma$ ($\Delta\rm TS = 13.2$) at energies  $>$ 100 TeV, respectively. These results confirm that the TeV source detected by LHAASO is indeed extended.  Using an elliptical gaussian intensity distribution instead of a Gaussian profile only marginally improves the fit likelihood by $1.6 \sigma$ ($\Delta\rm TS = 2.8$ ) at energies 8-100 TeV  and by a negligible amount at energies $> $100 TeV. This suggests that LHAASO cannot claim an obviously asymmetric structure as VERITAS has previously reported. To quantitatively compare the Gaussian and disk models, we evaluated their AIC values since these models are not nested. Due to a small difference in AIC ($\Delta\rm AIC < 4$), we are unable to clearly distinguish between these two extended models. For the subsequent analysis, we tentatively consider the Gaussian profile as our benchmark spatial model. With this choice, the significance is estimated to be  21$\sigma$ and 17$\sigma$ at 8$-$100 TeV and  $>$100 TeV, respectively, according to the TS values shown in Table~\ref{tab:SFR}.

\subsection{Spectrum}
\begin{figure}[ht]
  \renewcommand{\figurename}{Figure}
  \centering
  \includegraphics[scale=0.6]{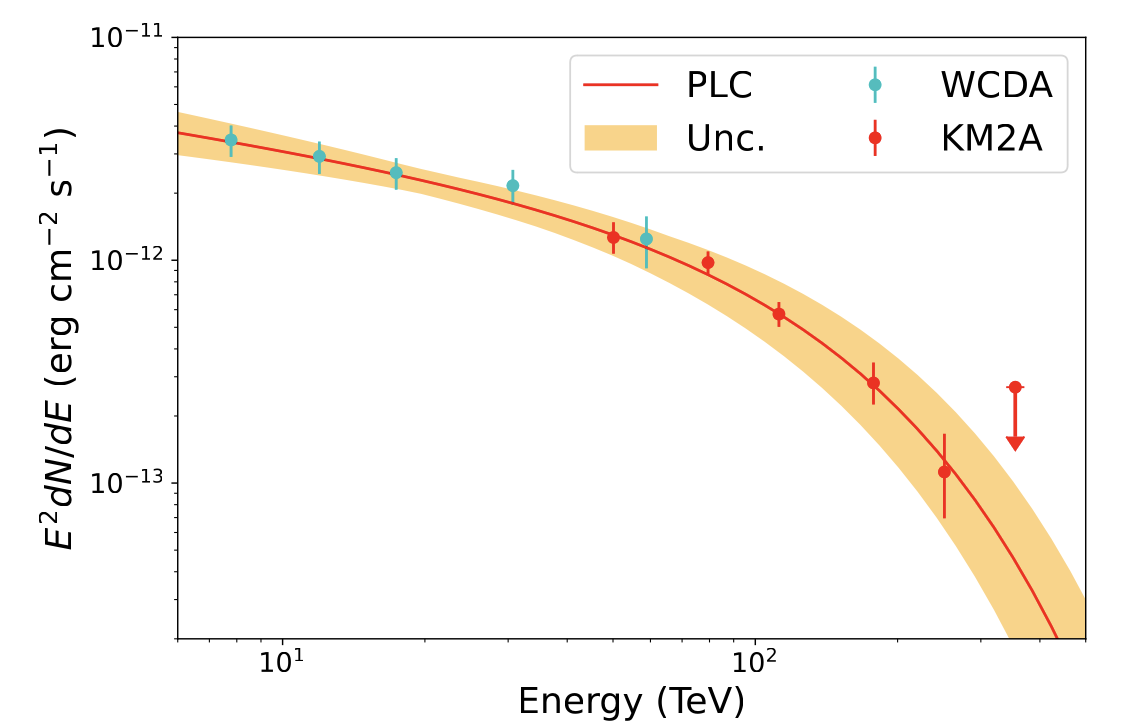}
 
  \caption{LHAASO differential $\gamma$-ray spectrum of the PWN CTA1. The red line and the orange butterfly represent the best-fit PLC model and its uncertainties, respectively. }
  
  \label{fig:sed_lhaaso}
\end{figure}

For the spectral analysis of CTA1, we performed a maximum likelihood fitting in the energy range above 8 TeV, using a Gaussian model described above. We compared three different spectral shapes for CTA1: a simple power-law (PL), a power-law with exponential cutoff (PLC), and a log-parabola (LP).  As shown in Table 2, the addition of a curvature in the spectrum is significantly improve the fit ($> 6\sigma$). Comparing the LP and PLC spectrum, the best spectral model is PLC with an improvement in modeling with $\Delta$ AIC = 8.4. Therefore, we can conclude that the best-fit  spectral model  is PLC model with an index of 2.31$\pm$0.13 and a cutoff energy of 110$\pm$25 TeV. The integral energy flux above 8 TeV is  $F_{> 8\rm\ TeV} \approx  4.9 \times 10^{-12 }\rm\ erg\ cm^{-2}\ s^{-1}$.  We determined the spectral points by performing a maximum likelihood analysis in each $N_{hit}$ or $E_{rec}$ bin considering the fixed PLC spectral shape. Upper limits on flux at a 95\% confidence level were derived when CTA1 had TS $<$ 4 (2$\sigma$) in a given bin.  The differential photon spectrum is shown in Figure ~\ref{fig:sed_lhaaso}, with spectral data points listed  in Table~\ref{tab:dfm}. 

\begin{table*}
 \centering
 \setlength{\tabcolsep}{0.08in}
 \caption{Spectral Fitting Results for CTA1 above 8 TeV \label{tab:SED}}
 \begin{tabular}{lcccccc}
 \hline
 \hline
\textbf{Model}&
\textbf{TS} &
\textbf{$N_0$} &
\textbf{$\alpha$}&
\textbf{$E_{0}$}&
\textbf{$E_{c}$}&
\textbf{AIC}\\
& &($10^{-16}$cm$^{-2}$ s$^{-1}$ TeV $^{-1}$) & &(TeV) &(TeV) & \\
\hline
    PL& 748.8 & 28.2 $\pm$ 1.5 & 2.81$\pm $0,04 & 20.0 & - & 0.0\\
    LP& 786.1 & 37.6 $\pm$ 3.2 &  2.48 $\pm$ 0.14 / 0.59 $\pm$ 0.16   & 20.0 & - & -35.3\\
    PLC&794.6 & 42.4$\pm$ 4.1 & 2.31$\pm$0.11 & 20.0 &110$\pm$25 & -43.7 \\
\hline
\hline
 \end{tabular}  
 \begin{tablenotes}
 \item Note: PL stands for the power-law model defined by $dN/dE=N_0 (E/E_0)^{- \alpha}$, PLC represents the power-law with exponential cutoff defined by $dN/dE=N_0 \times (E/E_0)^{- \alpha} exp[-(E/E_c)]$, and LP represents the log-parabola model defined by $F(E)=N_0 (E/E_0)^{-(\alpha+\beta log10(E/E_0))}$. For the LP model, the latter value listed in $\alpha$ column is the $\beta$ parameter. The AIC value is subtracted from that of the PL spectrum for a clear comparison.
 \end{tablenotes}
\end{table*}

Using the PLC spectrum presented in Table 2, the energy of the gamma-ray events from the CTA1 region were re-estimated event by event. We detailed the events with reconstruction energy above 250 TeV considering more stringent exclusion of cosmic ray background. Four gamma-like events with reconstructed energy larger than 250 TeV are listed in Table~\ref{tab:photons}. The maximum energy of the events is  $\approx 300\rm\ TeV$. 

\begin{table*}[h]
\setlength{\tabcolsep}{0.1in}
\caption{Differential Flux Measurements of CTA1 \label{tab:dfm}}
\doublerulesep 0.1pt 
\centering
\begin{tabular}{cccc} 
\toprule
\textbf{Detector}&
\textbf{Mid-Energy (TeV)} &
\textbf{Flux (cm$^{-2}$s$^{-1}$TeV$^{-1}$)} &
\textbf{TS}\\
\hline
 WCDA & $ 7.8 $ & $(3.58^{+0.57}_{-0.55}) \times 10^{-14}$ & 45.9 \\ 
         & $ 12.0 $ & $(1.27^{+0.21}_{-0.19}) \times 10^{-14}$ & 44.2\\
         &  $ 17.4 $ & $(5.11^{+0.83}_{-0.78}) \times 10^{-15}$ & 45.8 \\
         &  $ 30.7 $ & $(1.43 ^{+0.25} _{-0.22}) \times 10^{-15}$ &  58.9 \\ 
         & $ 58.8 $ & $(2.25 ^{+0.59} _{-0.57}) \times 10^{-16}$ & 20.8\\
    \hline
    KM2A &  $ 50.1 $ & $(3.53 ^{+0.57} _{-0.54}) \times 10^{-16}$ &  60.1\\
         &  $ 79.4 $ & $(9.17 ^{+0.82} _{-0.88} ) \times 10^{-17}$ & 191.3 \\
         &  $112.2 $ & $(2.95^{+0.37}_{-0.35}) \times 10^{-17}$ & 203.8 \\
         &  $177.8 $ & $(5.79^{+1.23}_{-1.10}) \times 10^{-18}$ & 97.8\\
         &  $251.1 $ & $(1.11^{+0.53}_{-0.41}) \times 10^{-18}$ & 20.2\\
         &  $354.8 $ & $<1.34 \times 10^{-18}$ & 1.3\\     
\hline
\end{tabular}
\begin{tablenotes}
\item Note: Mid-Energy represents the median energy corresponding to different bins. The error is statistic. The systematic uncertainties  of the flux are  estimated to 8\%. 
\end{tablenotes}
\end{table*}

\subsection{Systematic Uncertainties}
 To assess the robustness of our results we performed a number of systematic checks similar to those employed in \cite{2024ApJS..271...25C}.  The pointing error were estimated to $0.04^\circ$ for WCDA and KM2A data. A conservative estimation of the systematic error in Gaussian extension measurement was on the order of $0.08^\circ$. The systematic uncertainties in flux were estimated to be 8\%, which arise from the atmospheric model used in the Monte Carlo simulations. Furthermore, we assessed the uncertainties arising from the GDE by incorporating a dust template into our fitting. The impact on the flux was found to be less than  2\%.
\begin{table*}[h]
\caption{Photons with $E_{rec} > 250$ TeV for CTA1 \label{tab:photons}}
\centering
\doublerulesep 0.1pt 
\begin{tabular}{cccccccc}   
\toprule
\hline
\textbf{Index}&
\textbf{$E_{rec}$(TeV)}&
\textbf{$N_{e}$}&
\textbf{$N_{\mu}$}&
\textbf{$\Delta\theta$ (deg)} &
\textbf{Zenith angle (deg)}\\
\hline
1 & 309$_{-75}^{+100}$ &1024.3 & 1.8 &0.08 & 43.9\\
2 & 296$_{-91}^{+111}$ &1224.7 & 0.7 &0.36 &45.6\\
3 & 285$_{-85}^{+92}$ &1122.5 & 3.1 &0.34 &44.0\\
4 & 271$_{-70}^{+99}$ &1009.4 &1.2 &0.11 &44.9\\
\hline
\end{tabular}
\begin{tablenotes}
\item Note: $E_{rec}$ represents reconstructed energy based the PLC spectral shape. $N_e$ is the number of electromagnetic particles. $N_\mu$ is the number of muons detected in the region with a distance farther than 15 m from the core of the shower.
\end{tablenotes}
\end{table*}

\section{Discussion}

\subsection{The Nature of the $\gamma$-ray emission detected by LHAASO}

\begin{figure}[ht]
  \renewcommand{\figurename}{Figure}
  \centering
  \includegraphics[scale=0.4]{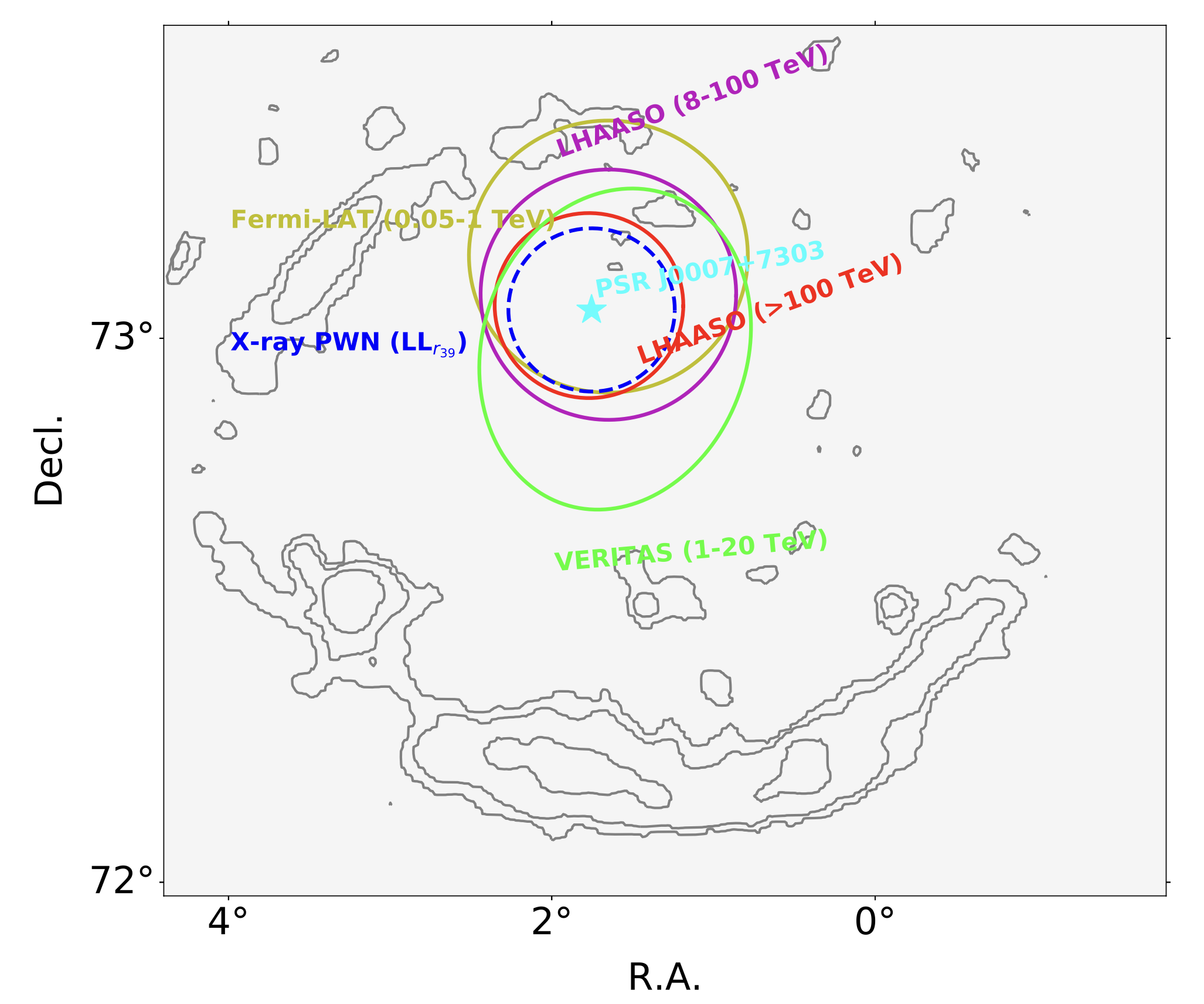}
  \caption{Multi-wavelength observations in CTA1 region. The circles represent 39\% flux region. The dashed  circle represents the lower limit of the extension. The size of the X-ray PWN has not been determined to date. The non-thermal X-ray emission is detected by ASCA in the $0.3^\circ$ region around the pulsar. Conservatively, the lower limit radius of the 39\% X-ray flux region is estimated to $0.15^\circ$, assuming a Gaussian flux distribution.}
  \label{fig:multi}
\end{figure}

To provide a clear identification of the $\gamma$-ray emission in the energy range from 8 TeV to $>$ 100 TeV detected by LHAASO, we compare its position and extension to observations in other bands, as shown in Figure~\ref{fig:multi}.  The $\gamma$-ray emission detected by LHAASO can be firmly associated with the X-ray PWN powered by the young pulsar PSR J0007+7303, as the position of the $\gamma$-ray emission coincides with that of X-ray PWN, while being clearly distinguished from the SNR radio shell. The non-thermal X-ray emission has an extension of approximately $0.3^\circ$ in ASCA observation~\citep{1997ApJ...485..221S} and $0.16^\circ$ in Suzaku observation~\citep{2012MNRAS.426.2283L}. Due to the constraints of the small field of view of the X-ray telescope and the influence of thermal X-rays, it is possible that the size of the X-ray PWN is more extensive. Conservatively, we consider the lower limit (LL) size of the X-ray PWN at $0.3^\circ$, and thus the 39\% flux radius of the X-ray PWN is larger than $0.15^\circ$ assuming a Gaussian X-ray flux distribution. There is no conflict in extension between the X-ray and $\gamma$-ray observations. We favor a one-zone leptonic model in which the same population of electrons contributes to both the X-ray and $\gamma$-ray emission. Additionally, evidence for leptonic process includes possible energy-dependent morphology detection. The cooling of energetic electrons as they are transported away from the pulsar results in the size of the emission shrinks and its center moves closer to the pulsar at higher energies. This behavior is supported by observations from Fermi-LAT, VERITAS, and LHAASO, with a confidence level of 2.5$\sigma$.

It is widely believed that the $\gamma$-ray radiation in the PWNe is produced in the leptonic scenario. However, as the IC emission above 100 TeV energies undergoes suppression due to the Klein-Nishina effect, the hadronic emission is possibly identified in UHE band. Indeed, Crab spectrum detected by LHAASO implies the possible contribution of the hadronic process due to the hardening at PeV energies\citep{2021ApJ...922..221L}. In the case of CTA1, we specifically investigate whether the UHE photons are dominated by the  $\pi^{0}$-decay gamma rays generated by the CR protons interacting with the interstellar medium (ISM). To investigate the environment around the pulsar, \cite{2016MNRAS.459.3868M} conducted a dedicated study based on a survey data of the intensity of CO line provided in \cite{2001ApJ...547..792D}. Most of the data in the velocity range $-26{\rm\ km\ s^{-1}} < v < -6\rm\ km\ s^{-1}$, which corresponds to the distance of CTA1, is tagged as noise. Due to the absence of a molecular cloud, we disfavor the UHE $\gamma$-ray emission originating from protons escaping a shock around CTA1 SNR and illuminating the molecular cloud. Additionally, specific regions were assigned a density of 0.017 and 0.037 $\rm\ cm^{-3}$ based on the X-ray absorption observations along the line of sight of CTA1, indicating a lower density  in the vicinity of CTA1 \citep{1997ApJ...485..221S}. Assuming an average density of $0.1\rm\ cm^{-3}$, which is the upper limit from the ISM density of the Galaxy in the direction of  CTA1, we can conservatively estimate a total energy of proton with energies $>$ 1 PeV to be $W_p > 1\times 10^{47} \rm \, ergs$ based on the UHE $\gamma$-ray flux observed. The energy budget is larger than the total energies from the pulsar spin-down power, roughly estimated by  $\dot{E}\tau_{c}\approx1\times10^{46}\rm\ ergs$. we can also exclude that the CTA1 PWN accelerates a sufficient number of PeV protons to dominate the UHE $\gamma$-ray emission observed.

\subsection{The broad-band spectrum: implications for the magnetic field and particle spectrum}

In the leptonic scenario, the broad-band nonthermal emission is dominated by two mechanisms: (1) synchrotron radiation for the emission from radio to X-ray, and (2) IC radiation for emission in the $\gamma$-ray band. In the one-zone model, the same population of relativistic electrons can produce the X-ray and $\gamma$-ray emission by interacting with the ambient magnetic field and the soft photon field, respectively. 

\begin{figure}[ht]
  \renewcommand{\figurename}{Figure}
  \centering

  \includegraphics[scale=0.6]{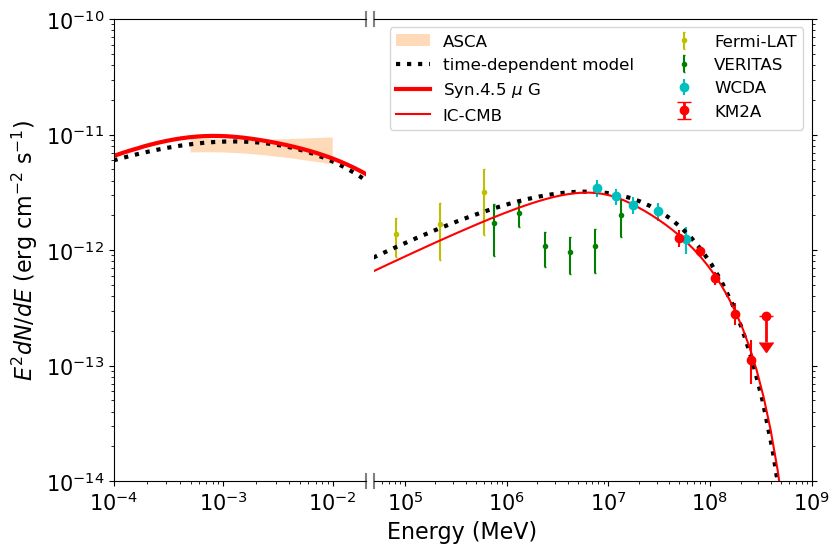}
 
  \caption{The differential energy spectrum of the PWN CTA1. The red line represents the expectation from a one-zone leptonic model, assuming a magnetic field strength of 4.5 $\mu G$ and considering only CMB target photons. The electron spectrum follows a broken power law distribution, where $dN_e/dE_e \propto E_e^{-3.13}\exp[-(E_e/373\rm\ TeV)^2]$ for energies $E> 40\rm\ TeV$  and $dN_e/dE_e \propto E_e^{-2.13}$ for energies $E< 40\rm\ TeV$. The time-dependent model (pure advection scenario) is referenced in the Appendix.}
  
  \label{fig:sed}
\end{figure}

The $\gamma$-ray emission is contributed by IC scattering of relativistic electrons interacting with several target photon fields, including the cosmic microwave background (CMB), infrared radiation field and star-light photon field. In the composite SNR CTA1 region, almost all of the VHE and UHE $\gamma$-ray fluxes come from the IC scattering with CMB, since the contributions from interactions with other photon fields are suppressed due to their low number densities and the Klein-Nishina effect \citep{2014JHEAp...1...31T,2016MNRAS.459.3868M}. 
Because the property of the CMB is well quantified, the $\gamma$-ray emission in LHAASO band could provide precise information about the spectrum of parent electrons. At energies above a few tens of TeV, the energy of the upscattered CMB photon $E_{\gamma}$ and that of the parent electron $E_{e}$ could be linked through the following simple relation \citep{2021Sci...373..425L}, 
\begin{equation}
    E_{e} \simeq 0.85 (E_{\gamma}/300\rm\ TeV)^{0.77}\rm\ PeV.
\end{equation}
The energy range of $\gamma$-rays detected by LHAASO is from $\approx 8 $ TeV to $\approx$ 300 TeV, which corresponds to the parent electron energy ranging from $\approx$ 50 TeV to $\approx$ 850 TeV. We assume that the electron spectrum follows a steady-state power-law distribution terminated by a super-exponential cutoff, i.e. $E_e^{-\alpha}\exp[-(E_e/E_{e,c})^2]$. The spectral index and cutoff energy of parent electron in the energy band  $50{\rm\ TeV} \lesssim E_{e}\lesssim 880\rm\ TeV$ can be constrained to $\alpha = 3.13\pm0.16$ and $E_{e,c} = 373\pm70$ TeV, respectively, by fitting to the LHAASO data as shown in Figure~\ref{fig:sed}.

The mean energies of the synchrotron ($E_{syn}$) and IC photons produced by the same population of electrons in the ambient magnetic field and CMB are related by  
\begin{equation}
    E_{syn} = 4.6 (E_{\gamma}/50{\rm\ TeV})^{1.5}(B/4.5\, \mu G) \rm\ keV.
\end{equation}
The energy range of the X-rays detected by ASCA is $E_{syn} = 0.5-10\rm\ KeV$, which corresponds to the $\gamma$-ray energy range of $E_{\gamma} \approx 11-83 \rm \ TeV$ considering a magnetic filed of 4.5 $\mu\rm G$. This implies that the observed X-ray and $\gamma$-ray emission from CTA1 are roughly produced by the population of electrons in the same energy range, in which the electron spectrum can be well constrained by LHAASO. Utilizing both the ASCA and LHAASO observations of the PWN, we can roughly constrain the current space-averaged magnetic field strength $B$ to $\sim$ 4.5 $\rm \mu G$ for the PWN CTA1, as shown in Figure~\ref{fig:sed}. 
The deduced magnetic field is at the same level with the interstellar magnetic field. Such low magnetic field has also been reported in other PWNe by analysing their synchrotron spectra, like HESS J1825-137 and HESS J1809-193. The magnetic field in young PWNe decreases with time during the free-expansion phase, due to the expansion of the volume of PWNe. Therefore, the magnetic field inside PWNe is expected to reach its minimum at the end of the free-expansion phase. We also note that although the magnetic field in PWN CTA1 is as low as the interstellar magnetic field, this does not necessarily imply that the turbulence is also similar. The possible stronger turbulence in the PWN could prevent particles from escaping to the ISM.
It's also worth noting that the non-thermal X-ray flux is not extracted from the entire TeV PWN region detected by LHAASO, but rather from a specific $0.3^\circ$ region within it. A more extended non-thermal X-ray survey around the entire TeV PWN is needed to provide a precise estimation of the current space-averaged magnetic field strength.

In the PWN CTA1, the energy loss of the electrons is primarily dominated by synchrotron radiation, with a cooling timescale given by $t_{c,syn}\approx 13.0 (E_e/40{\rm\ TeV})^{-1}(B/5 \mu G)^{-2} \rm\ kyr$. Considering a time-averaged magnetic field strength of   $B \sim 5\rm\ \mu G$, the cooling time roughly equals the pulsar's characteristic age of 13.9~kyr. Electrons with an energy greater than $\approx$ 40 TeV are expected to be in the fast cooling regime. Assuming that particle escape and the time evolution of energy injection rate can be neglected, the expected spectral index of the injected electrons can be approximated as $p \sim \alpha-1 = 2.13\pm0.16$ in the energy range from $\approx$ 40 TeV to $\approx 373\rm\, TeV$, considering the spectral steeping as being due to only the synchrotron loss. 
Conversely, electrons with $E_e \lesssim 40$ TeV are in the slow cooling regime, where the spectral index of the integrated electrons remains almost unchanged with respect to that of the injected spectrum. Therefore, a spectral break is expected in the steady-state electron spectrum at an energy around 40~TeV, which is indicated by the possible break observed in the $\gamma$-ray flux around the transition energy ($\sim$8~TeV) between the VERITAS and LHAASO energy range (see Figure~\ref{fig:sed}). 
In our time-dependent model, we take into account the decrease in energy injection rate and magnetic field strength over time, as well as the potential escape of particles. Through a detailed numerical simulation of the evolution of the particle distribution (see appendix), we find that an intrinsic injection spectrum with an index of $p=2.2$ could indeed result in a present electron spectrum with index of $\alpha \approx 3.11$ in  the energy range from $\approx$ 40 TeV to $\approx 350\rm\ TeV$, and of  $\alpha \approx 2.18$ below  $\approx$ 40 TeV. This is consistent with our previous simple explanation of the parent electron spectral properties of the PWN CTA1.

In the case of standard shock acceleration in the Bohm diffusion regime, a simple analytical presentation of the electron spectrum at shock over the entire energy range follows the form of a super-exponential cutoff, i.e., $Q(E_e)\propto E_e^{-p}\exp(-E_e^2/E_{e,max}^2)$~\citep{2007A&A...465..695Z}. 
Considering a compression ratio of 3-4, the index $p$ is estimated to be 2-2.25, which is consistent with the  spectral index of the injected electrons inferred by the LHAASO observations.
The maximum energy ($E_{e,max}$) of accelerated electrons is limited by the requirement of confinement of the particles inside the termination shock. This requires that the Larmor radius $R_L$ is smaller than the termination shock radius $R_s$, i.e., $R_L = \varepsilon R_s$, where $\varepsilon$ is the so-called containment factor with the value of $0<\varepsilon<1$. The Larmor radius of the electron with energy $E_{e}$ is 
\begin{equation}\label{rl}
    R_L = E_{e}/(e B_s),
\end{equation}
where $e$ is the electron charge, $B_s$ is the post-shock magnetic field strength, defined as \citep{1984ApJ...283..710K} 
\begin{equation}\label{bs}
    B_s \sim (3(\eta_B L(t)/c)^{0.5})/R_s,
\end{equation}
where $L(t)$ is the spin-down luminosity of the associated pulsar at time $t$, $\eta_B$ is the magnetic energy fraction. Using Equation (\ref{rl}) and (\ref{bs}) and the condition $R_L=\varepsilon R_s$, the maximum energy of the accelerated electrons is given by
\begin{equation}\label{emax}
    E_{\rm e,max} =3 e  \varepsilon \sqrt{\eta_{\rm B} L(t)/c}.
\end{equation}
 
We can see that the maximum energy of electrons evolves as the spin-down power with time. 
Taking into account  the spin-down power $\dot{E} = 4.5\times10^{35}\rm\ erg/s$ at current time, the maximum energy can be estimated as 
$E_{e,max} \approx 430 (\frac{\varepsilon}{0.16}) (\frac{\eta_B}{0.6})^{0.5} \rm\ TeV$. 
The cutoff energy of the steady-state electron spectrum, approximately 373 TeV, should be the time-integrated value over the last few thousand years. However, the electrons injected in earlier stages experienced more energy losses, the time-integrated average maximum energy is expected to be close to the maximum energy of the recently injected electrons, which indicates that the maximum energy of electrons injected at current time could be estimated by the maximum energy inferred from our observation. 
Moreover, in our time-dependent modeling, the cutoff energy of approximately 350 TeV in the accumulated electron spectrum (see Fig~\ref{fig:e_sed} in the Appendex) indeed roughly coincides with the maximum energy of 430 TeV in the currently injected electron spectrum. This implies that the cutoff energy derived from the steady-state electron spectrum could be a good estimation of the current particle acceleration ability of the termination shock in PWNe similar to CTA1.

\subsection{The Morphology: implications for propagation mechanism}

Relativistic particles accelerated at the termination shock propagate in PWN mainly by two transport mechanisms, i.e. advection and diffusion. To investigate the effects of particle advection and diffusion in the PWN, we performed detailed simulations of the spectrum and morphology of CTA1 for each transport scenario. The description of our model can be referred to the Appendix.

We first tune the parameters for both pure advection and pure diffusion scenarios to explain the multi-band non-thermal spectrum from the PWN CTA1. The parameters used to fit the flux data are almost similar in these two cases, except for those related to particle transport (see Table \ref{tab:para} in the Appendix). As a result, the calculated fluxes in the two cases are also similar, so we only plot the result in advection case for convenience (see Figure \ref{fig:sed}).

\begin{figure}[ht]
\centering
\includegraphics[width=0.7\textwidth]{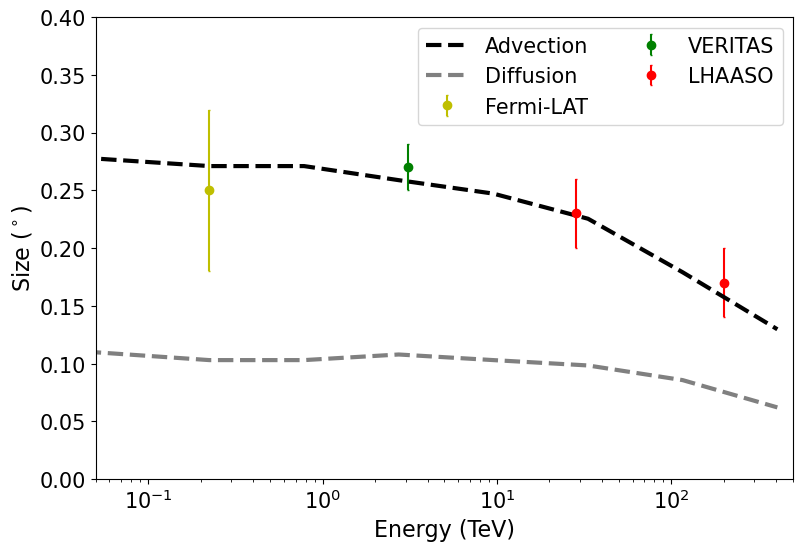}
\caption{The $\gamma$-ray extensions of the PWN CTA1. The simulated extension corresponds to the 39\% flux size, which means that the flux inside this region is 39\% of the total flux in the PWN. The black and grey dashed lines represent the extension calculated by the pure advection and pure diffusion scenarios.}
\label{fig:size}
\end{figure}

The morphology of the emission is an interplay between particle transport and energy losses in the PWN. Figure \ref{fig:size} shows the energy dependence of the $\gamma$-ray extension for different transport mechanisms. In our model, the $\gamma$-ray extension at a given energy refers to the 39\% flux size, in which the $\gamma$-ray flux is 39\% of the total flux in the PWN. It is clear that overall the extension of the $\gamma$-ray emission decreases with increasing photon energies in both pure advection and diffusion scenarios (for Kolmogorov turbulence), as the electrons responsible for producing higher energy $\gamma$-ray photons have a shorter lifetime, hence a shorter travelling distance. We can also see that the energy dependence of the size is not strictly a power-law in the whole energy range, because the electrons responsible for the lower energy $\gamma$-rays (below few TeV) are not completely cooled at present age. The calculated size in diffusion scenario is less dependent on energy than that in advection scenario (the curve is flatter), particularly in the higher energy range. This can be explained by the fact that higher energy electrons diffuse faster, which compensates the effect of stronger energy losses on the propagation distance.

Figure \ref{fig:size} shows that the size measured by Fermi-LAT, VERITAS and LHAASO could be well explained by the advection scenario, however, the extension calculated in the diffusion scenario is much smaller. The reason for this difference is that in the advection scenario, particles propagate faster in the inner nebula than in the outer region, causing them to accumulate at a larger radius. While in the case of diffusion, the particle density decreases roughly exponentially with increasing distance from the central source, resulting in a more compact distribution of particles. 
The different impacts of particle transport on particle distribution are illustrated in Figure~\ref{fig:Ne_dist} in the Appendix. For the diffusion scenario, it is unlikely to fit the data by simply increasing the diffusion coefficient, since larger diffusion coefficient means greater escape losses, which will make it hard to explain the observed flux. Moreover, when particles start to escape the PWN, further increasing diffusion coefficient would have very limited impact on the extension, because escaped particles are assumed not to contribute any emission outside the PWN. 
In fact, recent works suggest that escaped particles also produce IC emissions by interacting with the CMB photons, an effect which we did not consider here \citep{Martin_2024}. However, the similar size of the X-ray and $\gamma$-ray emission region may indicate that particle escape is not important for PWN CTA1, if the magnetic field is confined in the PWN.

In our simulation, the radius of PWN is a crucial factor which can affect the size of the emission. The PWN radius is mainly dependent on the mass of SN ejecta, the density of ISM, the initial spin-down luminosity and the age of the system. For simplicity, we assume that the evolution of PWN CTA1 is still in the free-expansion phase, and choose reasonable parameters to make the PWN radius as large as possible. For the chosen parameters (see Table~\ref{tab:para} in the Appendix), the PWN radius at present time is about 11.3~pc (A larger radius requires extreme parameters). The largest 39\% flux size of the $\gamma$-ray emission corresponding to this PWN radius in the pure diffusion scenario is around 0.1$^\circ$, which is significantly smaller than the observed size (see Figure~\ref{fig:size}). On the other hand, we can estimate the upper limit of the PWN radius by the fact that it should be smaller than the corresponding SNR radius, which is $\sim$ 20.4~pc infered from the radio observation \citep{2016MNRAS.459.3868M}. Even for such radius, the expected 39\% flux size in the diffusion scenario is $\sim$ 0.2$^\circ$, which still does not fit the data as well as the advection scenario. Therefore, the pure diffusion scenario is unlikely to explain the observed PWN size, and we suggest that the particle transport is dominated by advection in the nebula of CTA1.

The radius of PWN can also be influenced by its evolutionary stage. While we have assumed free-expansion phase in our calculation, it remains unclear whether the PWN CTA1 is in the free-expansion phase or the compression phase. Its relatively large age ($\sim$10~kyr) compared to the typical young PWNe indicates it may have already passed the free-expansion phase. Nonetheless, the large PWN radius ($\sim$10~pc indicated by observations) seems not to favor compression by the reverse shock. The attempt to fit the CTA1 spectrum using the reverberation model by \cite{2016MNRAS.459.3868M} also failed due to the incapability of fitting the VERITAS data. This may imply that the CTA1 is in an intermediate phase, where the PWN is currently under transition from the free-expansion phase to the compression phase. There is another possibility that some parts of the PWN are in free-expansion phase, while other parts in compression phase, due to different density of the ISM in the northwest and southeast direction to the PWN, as suggested by \cite{2016MNRAS.459.3868M}. If the PWN CTA1 is indeed in this intermediate phase, our modelling based on the assumption of free-expansion phase may still applies to some extent, since the compression effect of the reverse shock may be relatively small at that time. Finally, a more sophisticated PWN model is needed to study particle transport during the reverberation phase. Additionally, future detailed observations in radio and X-rays may help distinguish between the different evolutionary stages of the PWN CTA1.

\section{Conclusions}
Using about three years of KM2A data and two years of WCDA data, we have a deep view of  $\gamma$-rays emission from composite SNR CTA1 in energy range from $\approx$8 TeV to $\approx$ 300 TeV.  The LHAASO source is detected with a significance of $21\sigma$  in the energy band 8-100 TeV and  $17\sigma$ at energies above 100 TeV.  The source is a significant extended source with a 39\% flux radius of $\approx 0.23^\circ$ at energy band 8-100 TeV and of $\approx 0.17^\circ$ at energies above 100 TeV, based on a  Gaussian profile.  The best spectral model is PLC with $dN/dE = (42.4\pm4.1)(\frac{E}{20\rm\ TeV})^{-2.31\pm0.11}\exp(-\frac{E}{110\pm25\rm\ TeV})$ $\rm\ TeV^{-1}\ cm^{-2}\ s^{-1}$ in the energy range from $\approx$ 8 TeV to $\approx$ 300 TeV.  The integral energy flux above 8 TeV, $F_{> 8\rm\ TeV} \approx  4.9 \times 10^{-12 }\rm\ erg\ cm^{-2}\ s^{-1}$.   Given that the emission is extended, with a centroid near the pulsar PSR J0007+7303 and its X-ray PWN, we confirm that the $\gamma$-ray emission detected by LHAASO is physically associated with the PWN in the composite SNR CTA1. Additionally, the plausible morphological evolution in the $\gamma$-ray emission can be confirmed by combining Fermi-LAT, VERITAS and LHAASO observations, further implying the $\gamma$-ray leptonic origin. We also investigate whether or not the UHE $\gamma$-ray emission is from a hadronic process. Due to the absence of a molecular cloud and the lower density of the ISM, we can almost rule out the UHE $\gamma$-ray emission being dominated by a hadronic process. In the leptonic scenario, we can accurately determine the electron spectrum above 50 TeV energies. The steady-state electron spectrum index of approximately 3.13 implies that the intrinsic injected electrons index should be around 2.13, which is consistent with the expected acceleration mechanism at the terminal shock of PWNe. The cutoff energy might be treated as a good indicator of the maximum energy of the current injected spectrum. Combined with X-ray observations, the current space-averaged magnetic field of the CTA1 PWN can be roughly limited to $\sim 4.5\rm\ \mu G$. Under the assumption of free-expansion phase, we favor advection dominating the particle transportation in the CTA1 PWN, in order to satisfy the multi-wavelength spectrum and the $\gamma$-ray extensions.

\section*{Acknowledgements}

We would like to thank all staff members who work at the LHAASO site above 4400 meters above sea level year-round to maintain the detector and keep the water recycling system, electricity power supply and other components of the experiment operating smoothly. We are grateful to Chengdu Management Committee of Tianfu New Area for the constant financial support for research with LHAASO data. We deeply appreciate the computing and data service support provided by the National High Energy Physics Data Center for the data analysis in this paper. This research work is also supported by the following grants: in China by the National Natural Science Foundation of China NSFC No. 12393851,  No.12393854, No.12393852, No.12393853, No.12022502, No.12205314, No.12105301, No.12261160362, No.12105294, No.U1931201, No.2024NSFJQ0060, and in Thailand by the National Science and Technology Development Agency (NSTDA) and the National Research Council of Thailand (NRCT) under the High-Potential Research Team Grant Program (N42A650868)..

\section*{AUTHOR CONTRIBUTIONS}
S.Q. Xi, S.Z. Chen and B. Li led the drafting of the text. Y.Z. Li and S.Q. Xi conducted the data analysis, and Y. Huang performed the cross-check for the data analysis. B. Li and S.Q. Xi led the modeling efforts. S.Z. Chen and S.C. Hu provided the sky map and contributed to the background estimation. Zhen Cao, the spokesperson of the LHAASO Collaboration and the principal investigator of the LHAASO project. All other authors participated in data analysis, including detector calibration, data processing, event reconstruction, data quality checks, and various simulations, and provided valuable comments on the manuscript.

\section*{Appendix}

\setcounter{equation}{0}
\setcounter{table}{0}
\setcounter{figure}{0}
\renewcommand{\theequation}{A\arabic{equation}}
\renewcommand{\thetable}{A\arabic{table}}
\renewcommand{\thefigure}{A\arabic{figure}}
\renewcommand*{\theHequation}{\theequation}
\renewcommand*{\theHtable}{\thetable}
\renewcommand*{\theHfigure}{\thefigure}

To obtain the particle spatial and energy distribution in PWN, we need to solve the advection-diffusion equation (assuming spherical symmetry):

\begin{equation} \label{eq:adv-diff}
    \frac{\partial N_{\rm{e}}(r, E_{\rm e}, t)}{\partial t} = Q_{\rm{e}}(E_{\rm e}, t) + \frac{1}{r^2} \frac{\partial}{\partial r} \left[r^2 D(E_{\rm{e}},t) \frac{\partial N_{\rm e}(r, E_{\rm e}, t)}{\partial r} \right] - \frac{1}{r^2} \frac{\partial}{\partial r}\left[r^2 V(r)N_{\rm{e}}(r, E_{\rm e}, t)\right] + \frac{\partial}{\partial E_{\rm{e}}} \left[\dot{E}_{\rm{e}} N_{\rm{e}}(r, E_{\rm e}, t)\right],
\end{equation}
where $N_{\rm{e}}(r, E_{\rm e}, t)$ is the differential number density of electrons and positrons. The first term of the equation is the source term. The injection spectrum of electrons is assumed to follow a broken power-law with supper exponential cutoff distribution. It's a good approximation to express the spectrum as  \citep{2021Sci...373..425L}
\begin{equation}
    Q_{\rm e}(E_{\rm e}, t)=Q_{0}(t) E_{\rm e}^{-\alpha_2}\left[1+\left(E_{\rm e} / E_{\rm b}\right)^{-(\alpha_2-\alpha_1)}\right]^{-1} \exp \left[-\left(E_{\rm e} / E_{\rm e, max}(t)\right)^{2}\right], 
\end{equation}
where $Q_0(t)$ is the normalization factor, which can be determined by solving $L(t)=\int_{E_{\rm e, min}}^{E_{\rm e, max}} E_{\rm e}Q(E_{\rm e},t) \, {\rm d} E_{\rm e}$. $E_{\rm e, min}$ and $E_{\rm e, max}$ are the minimum and maximum energy of injected electrons. The maximum energy of electrons accelerated at the termination shock could be estimated by Equation (\ref{emax}). $E_{\rm b}$ is the break energy. $\alpha_1$ and $\alpha_2$ are the spectral indexes of the lower energy and higher energy part of the electron spectrum.

Assuming a braking index $n=3$, the spin-down luminosity $L(t)$ of the pulsar evolves with time as \citep{2006ARA&A..44...17G}
\begin{equation}
    L(t)=\dot{E} \frac{\left(1+t_{\text {age }} / \tau_{0}\right)^{2}}{\left(1+t / \tau_{0}\right)^{2}},
\end{equation}
where $\dot{E}$ is the pulsar's spin-down power at its current age $t_{\rm age}$
\begin{equation}
t_{\rm age}=\tau_{\rm c}\left[1-(\frac{P_0}{P})^2\right],
\end{equation}
with $\tau_{\rm c}=P/2\dot{P}$ the characteristic age of pulsar. $P_0$ and $P$ are the initial and current spin periods, respectively. The initial spin-down timescale of the pulsar is defined as 
\begin{equation}
    \tau_0=\tau_{\rm c} (\frac{P_0}{P})^2.
\end{equation}

The second term on the right hand side of Equation (\ref{eq:adv-diff}) is the diffusion term. We assume that the diffusion coefficient is homogeneous in the PWN, while it has time and energy dependence of $D(E_{\rm{e}},t)=D_{0}(t)\left(E_{\rm{e}} / 1 \, \rm{TeV}\right)^{\delta}$, with $D_0(t)$ being the normalization factor at 1~TeV and $\delta=1/3$ for the Kolmogorov turbulence.
$D_0(t)$ is assumed to be inversely proportional to the magnetic field $B(t)$.
The evolution of magnetic field $B(t)$ in PWN could be calculated by \citep{1973ApJ...186..249P}
\begin{equation} \label{eq:B-field}
    \frac{d W_{\mathrm{B}}(t)}{d t}=\eta_{\mathrm{B}} L(t)-\frac{W_{\mathrm{B}}(t)}{R_{\mathrm{pwn}}(t)} \frac{d R_{\mathrm{pwn}}(t)}{d t},
\end{equation}
where $W_{\rm B}=B^2 R^3_{\rm pwn}/6$ is the total magnetic energy in PWN and $\eta_{\rm B}$ the fraction of the spin-down luminosity converted into magnetic energy. 

The third term describes particle advection by the flow downstream the termination shock. The bulk velocity of the downstream plasma flow is supposed to decreases with radius as $V(r)=V_0 (r/R_{\rm ts})^{-\beta}$, where $R_{\rm ts}(t)$ is the radius of termination shock and $\beta$ ($0\leq \beta \leq 2$) describes the spatial dependence of advection velocity.

The last term represents the energy losses of relativistic electrons due to synchrotron radiation, IC scattering and adiabatic cooling, which are given by
\begin{equation} \label{eq:loss}
    \dot{E}_{\rm{e}} = \frac{4}{3} \sigma_{\rm T} \gamma^{2}\left[U_{\rm B}+\sum \frac{U_{i}}{\left(1+4 \gamma \epsilon_{0, i}\right)^{3 / 2}}\right]+\frac{E_{\rm e}}{3} \frac{1}{r^2} \frac{\partial}{\partial r} \left[r^2 V(r)\right],
\end{equation}
where $U_{\rm B}=B^2/8\pi$ is the energy density of magnetic field and $U_i$ is the energy density of the $i$-th component of the interstellar radiation field (ISRF). 
Following \cite{Torres_2014}, we consider three components for the radiation field in the direction of CTA1: the cosmic microwave background (CMB, $T$=2.73~K, $U$=0.25~eV~cm$^{-3}$), far-infrared radiation field ($T$=70~K, $U$=0.1~eV~cm$^{-3}$), and near-infrared radiation field ($T$=5000~K, $U$=0.1~eV~cm$^{-3}$). $\epsilon_{0, i}$ is the average photon energy of the radiation field, which equals to $2.8k_{\rm B}T_i$ for blackbody/greybody radiation with $T_i$ being the temperature.

Equation (\ref{eq:adv-diff}) is solved numerically in this work. The computational domain is the region between the termination shock and the outer boundary of the PWN. The evolution of the radii of termination shock and PWN ($R_{\rm pwn}(t)$ and $R_{\rm ts}(t)$) could be obtained using the method given by \cite{Gelfand_2009}. At the termination shock, the inner boundary condition requires that the number of particles injected from the upper stream should equal to that transported downstream by advection and diffusion \citep{Peng_2022}. At the outer boundary, a free escape condition is imposed to simulate the particle escape from PWN. The model described above is then applied to the nebula of CTA1. The parameters of the model used to simulate the the particle distribution and emission from the nebula of CTA1 are shown in Table~\ref{tab:para}.

\begin{table}[ht]
    \begin{center}
        \caption{Parameters used to fit the SED and $\gamma$-ray extensions of the PWN CTA1 in the pure advection and pure diffusion scenarios.}
        \label{tab:para}
        \begin{tabular}{llcc}
            \hline
            \hline
            Measured parameter & Symbol & \multicolumn{2}{c}{Value} \\
            \hline
            Spin period (s) & $P$ & \multicolumn{2}{c}{0.316} \\
            Period derivative (s~s$^{-1}$) & $\dot P$ & \multicolumn{2}{c}{$3.6 \times 10^{-13}$} \\
            Characteristic age (kyr) & $\tau_{\rm c} = P/2\dot P$ & \multicolumn{2}{c}{13.9} \\
            Spin-down luminosity (erg~s$^{-1}$) & $\dot E$ & \multicolumn{2}{c}{$4.5\times 10^{35}$} \\
            Distance to pulsar (kpc) & $d$  & \multicolumn{2}{c}{1.4} \\
            \hline
            Hypothetical parameter &  &  & \\
            \hline
            SN explosion energy (erg) & $E_{\rm SN}$ & \multicolumn{2}{c}{$10^{51}$} \\
            Ejected mass ($M_{\odot}$) & $M_{\rm ej}$ & \multicolumn{2}{c}{20.0} \\
            ISM density (cm$^{-3}$) & $n_{\rm ISM}$ & \multicolumn{2}{c}{0.1} \\ 
            Braking index & $n$ & \multicolumn{2}{c}{3.0} \\
            Initial spin period (s) & $P_0$ & \multicolumn{2}{c}{0.12} \\
            Initial spin-down timescale (kyr) & $\tau_0$ & \multicolumn{2}{c}{2.0} \\
            Initial spin luminosity (erg~s$^{-1}$)& $\dot E_0$ & \multicolumn{2}{c}{$2.16\times 10^{37}$}\\
            Electron minimum energy (MeV) & $E_{\rm e, min}$ & \multicolumn{2}{c}{1} \\
            Electron break energy (TeV) & $E_{\rm b}$ &\multicolumn{2}{c}{0.05} \\
            \hline
            Fitted parameter (at present time) &  & \it{Advection} & \it{Diffusion} \\
            \hline
            SNR radius (pc) & $R_{\rm snr}$ & 20.6 & 20.6 \\
            TS radius (pc) & $R_{\rm ts}$ & 0.6 & 0.6 \\
            PWN radius (pc) & $R_{\rm pwn}$ & 11.3 & 11.3 \\
            Conversion efficiency & $\eta_{\rm e}$ & 0.4 & 0.4 \\
            Magnetic efficiency & $\eta_{\rm B} $ & 0.6 & 0.6 \\
            Spectral index & $\alpha_1$ & 1.5 & 1.5 \\
                                    & $\alpha_2$ & 2.13 & 2.1 \\
            Containment factor & $\epsilon$ & 0.16 & 0.15 \\
            Current magnetic field ($\mu$G) & $B$ & 4.2 & 4.2 \\
            Advection velocity ($c$) & $V_0$ & 0.9 & - \\
              & $\beta$ & 1.9 & - \\
            Diffusion coefficient (cm$^2$~s$^{-1}$) & $D_0$ & - & $1.92\times 10^{26}$ \\
            \hline
        \end{tabular}
    \end{center}
\end{table}

\begin{figure}[ht]
\centering
\includegraphics[width=0.4\textwidth]{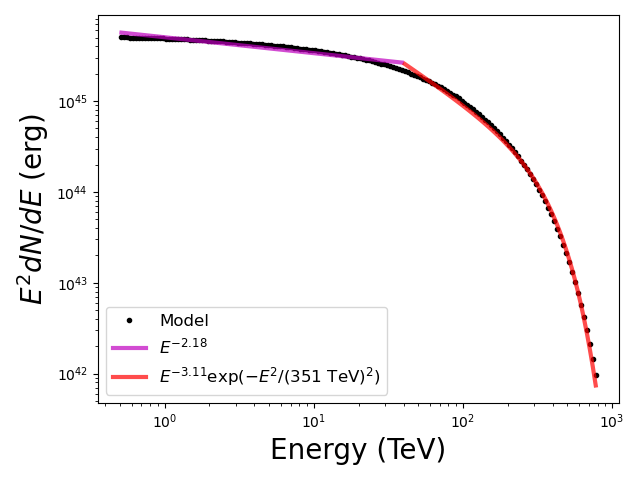}
\caption{The spectral energy distribution of the electrons in PWN CTA1.}
\label{fig:e_sed}
\end{figure}

\begin{figure}[ht]
\centering
\includegraphics[width=\textwidth]{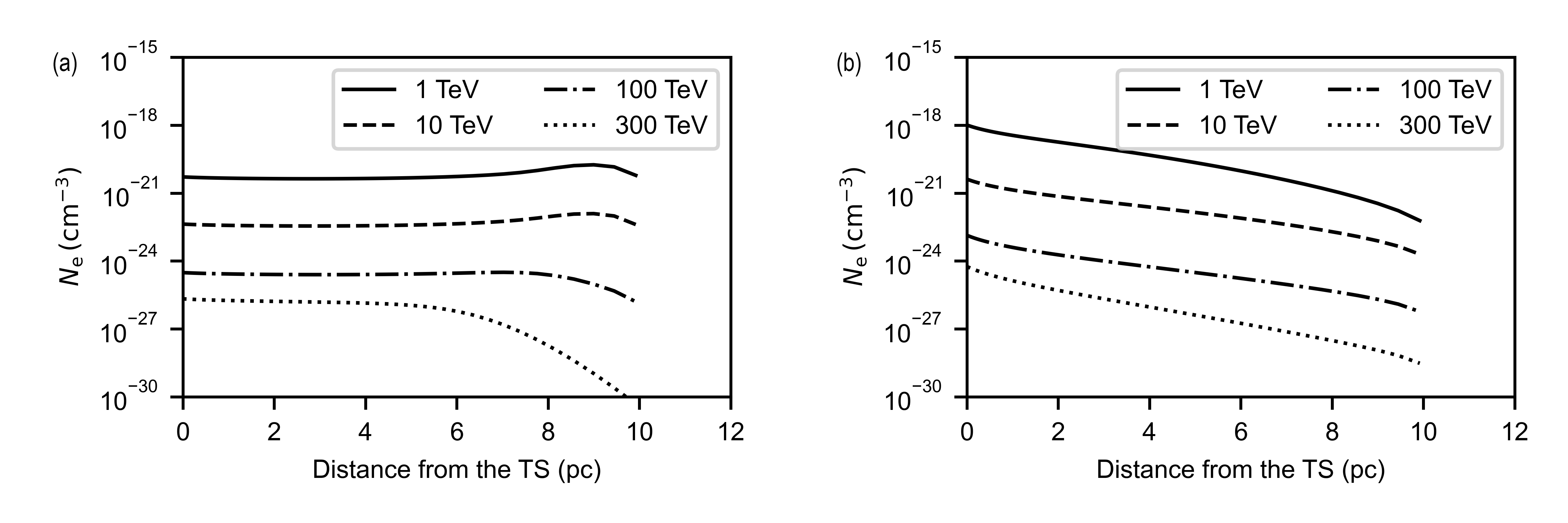}
\caption{\textbf{Particle number densities as a function of the distance from the termination shock in different transport scenarios. The parameters used to plot are given in Table~\ref{tab:para}. The particle distribution for the advection case (a) is nearly constant for the flow velocity profile $V \propto r^{-1.9}$. The bump near the PWN outer boundary is due to the higher pulsar spin-down power in the early times. In the diffusion case (b), the particle distribution decreases nearly exponentially with the radius.}}
\label{fig:Ne_dist}
\end{figure}

The time-integrated electron spectrum calculated by our model is shown in Figure~\ref{fig:e_sed}. In order to compare the numerical result with our measurements, we use a broken power-law with an exponential cutoff function to fit the spectrum from 0.5 TeV to 1 PeV, in which the break energy is fixed at 40 TeV. The spatial distribution of electrons in different transport scenarios are plotted in Figure~\ref{fig:Ne_dist}.

\bibliography{reference}
\bibliographystyle{aasjournal}


\clearpage
\input{Science20251226}

\end{document}

%% file: Science20251226.tex
Zhen Cao$^{1,2,3}$,
F. Aharonian$^{3,4,5,6}$,
Y.X. Bai$^{1,3}$,
Y.W. Bao$^{7}$,
D. Bastieri$^{8}$,
X.J. Bi$^{1,2,3}$,
Y.J. Bi$^{1,3}$,
W. Bian$^{7}$,
J. Blunier$^{9}$,
A.V. Bukevich$^{10}$
% ...
\\
$^{1}$ State Key Laboratory of Particle Astrophysics \& Experimental Physics Division \& Computing Center, Institute of High Energy Physics, Chinese Academy of Sciences, 100049 Beijing, China\\
$^{2}$ University of Chinese Academy of Sciences, 100049 Beijing, China\\
$^{3}$ TIANFU Cosmic Ray Research Center, 610000 Chengdu, Sichuan,  China\\
$^{4}$ University of Science and Technology of China, 230026 Hefei, Anhui, China\\
$^{5}$ Yerevan State University, 1 Alek Manukyan Street, Yerevan 0025, Armeni a\\
$^{6}$ Max-Planck-Institut for Nuclear Physics, P.O. Box 103980, 69029  Heidelberg, Germany\\
$^{7}$ Tsung-Dao Lee Institute \& School of Physics and Astronomy, Shanghai Jiao Tong University, 200240 Shanghai, China\\
$^{8}$ Center for Astrophysics, Guangzhou University, 510006 Guangzhou, Guangdong, China\\
$^{9}$ APC, Universit\'e Paris Cit\'e, CNRS/IN2P3, CEA/IRFU, Observatoire de Paris, 119 75205 Paris, France\\
$^{10}$ Institute for Nuclear Research of Russian Academy of Sciences, 117312 Moscow, Russia\\
% ...
